\pdfoutput=1
\documentclass{article}
\usepackage{caption} 
\usepackage{graphicx}
\usepackage{amsmath}
\usepackage{geometry}
\usepackage{authblk}
\usepackage{hyperref}
\usepackage[T1]{fontenc}
\usepackage[utf8]{inputenc}

\title{Entanglement Classifier in Chemical Reactions}
\author{Junxu Li and Sabre Kais}
\affil{Department of Chemistry, Department of Physics and Astronomy, and Birck Nanotechnology Center,
	Purdue University, West Lafayette, IN 47907, United States}

\begin{document}
	\maketitle
	
	\begin{abstract}
		
		Ever since the appearance of the seminal work of Einstein, Podolsky and Rosen (The EPR-paradox), the phenomenon of entanglement, which features the essential difference between classical and quantum physics, has received wide theoretical and experimental attentions.  Recently the desire to understand and create  quantum entanglement between particles such as spins, photons, atoms and molecules is fueled by the development of quantum teleportation, quantum communication, quantum cryptography and quantum computation. Although most of the work has focused on showing that entanglement violates the famous Bell’s inequality and its generalization for discrete measurements, few recent attempts focus on continuous measurement results. Here, we have developed a general practical inequality to test entanglement for continuous measurement results, particularly scattering of chemical reactions. After we explain how to implement this new inequality to classify entanglement in scattering experiments, we propose a specific chemical reaction to test the violation of this inequality. The method is general and could be used to classify entanglement for continuous measurement results.

	\end{abstract}
	
	\section*{Introduction}
	
	Entanglement, which is a quantum mechanical property that describes a correlation between quantum mechanical systems that has no classical analog  was introduced by  Shr\"odinger in 1935 \cite{schrodinger1935gegenwartige}. Such phenomenon was the subject of the famous paper by Einstein, Podolsky, and  Rosen, known as the EPR paradox\cite{einstein1935can}, where they considered such behavior to be impossible and argued that the accepted formulation of quantum mechanics must therefore be incomplete. The debate lasted for nearly thirty years until the proposal of Bell's inequality which is violated by entanglement\cite{bell1964einstein}.  
	Entanglement effect was verified experimentally\cite{aspect1981experimental} in tests where the polarization of photons and  spins of entangled particles were measured 
	to be statistically violating Bell's inequality.

	Nowadays entanglement has become an extremely important physical resource for many applications in quantum communication\cite{ursin2007entanglement,zhang2017quantum,mazeas2016high,kurpiers2018deterministic} and quantum computation\cite{sorensen2000entanglement,sheng2015deterministic,o2007optical,jozsa2003role,ekert1998quantum,wei2016quantum,kais2007entanglement}.

	Tow-particle entanglement has long been demonstrated experimentally and recently there are significant achievements to generate entanglement between three and more spatially separated particle systems. For instance, ten-photon entanglement system was successfully generated experimentally in 2016 \cite{reimer2016generation, wang2016experimental}. There are also observations of entanglement in quantum dots\cite{bayer2001coupling}, NV centers in diamond\cite{neumann2008multipartite}, trapped ions\cite{molmer1999multiparticle} and even entanglement between photon and quantum dots\cite{gao2012observation}.  
	
	It is very important in all of these experiments to be able to quantify or measure the entanglement. Several methods have been proposed to address this question such as entanglement witnesses\cite{toth2005detecting}, entropic inequalities\cite{cerf1997entropic}, and  quantum states classifier based on machine learning\cite{gao2018experimental}. However, most of these methods are designed for discrete measurement results.  Although there are a few successful theoretical analysis for continuous measurement results, they have focused  mainly on photonic systems\cite{wenger2003maximal,chen2002maximal,he2010bell}. The most widely used method to classify entanglement is quantum tomography, by which one could obtain the density matrix of the system\cite{d2001quantum} from experimental measurements. However, quantum tomography is much time and resource consuming
	as it scales exponentially with the system size\cite{altepeter20044}.

	In this paper, we propose a general practical method to classify entanglement for continuous measurement results. We introduce auxiliary functions to simplify the complicated measurement results and develop a generalized Bell’s type inequality for continuous measurements. We also propose an experimental design to test the method in scattering of chemical reactions. Here, we designed a practicable experiment based on the recent scattering experiments of the oriented molecule HD and H$_2$ molecules\cite{perreault2018cold}. Based on recent Zare and coworker experimental data\cite{perreault2018cold}, we simulate the possible measurement results and demonstrate how to distinguish entangled states from un-entangled ones. Moreover, our work also provides the possibility to classify entanglement in other chemical reactions as suggested by Brumer and coworkers for entanglement assisted coherent control in chemical reactions \cite{zeman2004coherent,gong2003entanglement}.

	\section{Bell’s inequality for continuous measurement results}

	We start by preparing N-particles in a pure state 
	$|\Phi\rangle$. If we perform an $r$-th measurement on them, they will collapse on the eigenstate set $\{|\phi_r^1\rangle, |\phi_r^2\rangle, \cdots ,|\phi_r^n\rangle\}$. Then we could expand $|\Phi\rangle$ as
	\begin{equation}
	|\Phi\rangle=\sum_{i=1}^{n}{\alpha_r^i|\phi_r^i\rangle}
	\end{equation}
	where $|\phi_r^i\rangle$ is the $i$-th eigenstate corresponding to the measurement $r$, and $|\Phi\rangle$ is normalized, $\sum_{i}^{n}{|\alpha_r^i|^2}=1$. Performing measurement $r$, we can obtain the measurement results (denoted here by spectrum)  $S(|\Phi\rangle\langle\Phi|, r, \bf{x})$ where $(|\Phi\rangle\langle\Phi|$ is the density matrix, $r$ represents the measurement $r$ and the variable ${\bf x}$ could be, for instance, the scattering angle in scattering experiment.

	The distribution of the measurement results (spectrum) can be written as
	\begin{equation}
	S(|\Phi\rangle\langle\Phi|, r, {\bf x}) = \sum_{i}^{n} |\alpha_r^i|^2 \cdot S(|\phi_r^i\rangle\langle\phi_r^i|, r, {\bf x})
	\label{discription}
	\end{equation}
	where $S(|\phi_r^i\rangle\langle\phi_r^i|,r, \bf{x})$ is the spectrum of the state $|\phi_r^i\rangle$ under the $r$-th  measurement. 
	
	In the case of the spin-system to be discussed in the supplementary material, a quite common example satisfying Eq.(\ref{discription}) is the measurement in Stern-Gerlach (SG) experiment\cite{gerlach1922experimentelle} of spin $\frac{1}{2}$ particles. If the particles are prepared in the pure  state
	\begin{equation*}
	|\Phi\rangle=\alpha_z^+|\uparrow\rangle+\alpha_z^-|\downarrow\rangle
	\end{equation*}
	Then providing enough particles to go through that apparatus, the final spectrum can be written as 
	\begin{equation*}
	S(|\Phi\rangle\langle\Phi|,r, {\bf x}) = |\alpha_z^+|^2\cdot S(|\uparrow\rangle\langle\uparrow|,z, {\bf x})+
	|\alpha_z^-|^2\cdot S(|\downarrow\rangle\langle\downarrow|,z, {\bf x})
	\end{equation*}
	Here $z$ represents the direction of the magnetic field in SG apparatus, $S(|\uparrow\rangle\langle\uparrow|,z, {\bf x})$ and $S(|\downarrow\rangle\langle\downarrow|,z, {\bf x})$ describes the distribution of particles detected at different location,$ {\bf x}$,  on the screen for particles at state $|\uparrow\ \rangle$ and $|\downarrow\ \rangle$ respectively. 	
	
	If these particles are prepared as a mixed state whose density matrix is given by:
	\begin{equation}
	\rho_1=\sum_{i=1}^{m}p_i|\Phi_i\rangle\langle \Phi_i|
	\end{equation}
	where $\sum_{i=1}^{m}p_i=1$. The spectrum under $r$-th  measurement is given by  
	\begin{equation}
	S(\rho_1,r, {\bf x})=\sum_{i=1}^{m}p_iS(|\Phi_i\rangle\langle \Phi_i|,r, {\bf x})
	\label{discription2}
	\end{equation}
	In the following section, we will assume that Eq.(\ref{discription}) and Eq.(\ref{discription2}) are always satisfied,
	and we will mainly focus on 2-particle systems. Generally, if we use $|\Psi\rangle$ to represent a pure state of the 2-particle system, its density matrix can be written as:
	\begin{equation}
	\rho=\sum_{i=1}^{m}p_i|\Psi_i\rangle\langle \Psi_i|
	\label{densityMatrix}
	\end{equation}
	with 
	\begin{equation}
	|\Psi_i\rangle=\sum_{j_1,j_2}c_i(k_1,k_2,j_1,j_2)|\phi_{k_1}^{j_1}\phi_{k_2}^{j_2}\rangle
	\label{state}
	\end{equation}
	where $\sum_{i=1}^{m}p_i=1$, and $\sum_{j_1,j_2}|c_i(k_1,k_2,j_1,j_2)|^2=1$. Here $|\phi_{k_1}^{j_1}\rangle$ is the $j_1$-th eigenstate of a single particle under the $k_1$-th measurement, and $|\Psi\rangle$ is expanded in the basis set $\{|\phi_{k_1}^{j_1}\phi_{k_2}^{j_2}\rangle\}$. In this section, for simplicity, we only consider the situation that a single particle have just 2 eigenstates $|\phi_{r}^{+}\rangle$ and $|\phi_{r}^{-}\rangle$ under $r$-th  measurement. Positive partial transpose (PPT) criterion (or Peres–Horodecki criterion)\cite{peres1996separability} offers us a method to measure entanglement in such a 2-particle system: If the partial transpose of the density matrix $\rho^{T_B}$ has any non-negative eigenvalue, then $\rho$ is entangled\cite{peres1996separability, rudolph2004computable}.

	Now,  we can rewrite Eq.(\ref{state}) as:
	\begin{equation}
	\begin{split}
	|\Psi_i\rangle=&c_i(r,t,++)|\phi_{r}^{+}\phi_{t}^{+}\rangle + c_i(r,t,+-)|\phi_{r}^{+}\phi_{t}^{-}\rangle
	\\
	+&c_i(r,t,-+)|\phi_{r}^{-}\phi_{t}^{+}\rangle + c_i(r,t,--)|\phi_{r}^{-}\phi_{t}^{-}\rangle
	\end{split}
	\end{equation}
	

	Suppose that these particles are divided into two channels, and the $r$-th  and $t$-th  measurements are carried on for channel 1 and 2 respectively, one can obtain the spectrum based on Eq.(\ref{discription}) as:
	\begin{equation}
	\begin{split}
	S(\rho, r, t, {\bf x_1}, {\bf x_2})= 
	&\sum_{i=1}^{m}{p_i|c_i(r,t,++)|^2 }\cdot S(|\phi_{r}^+\rangle\langle \phi_{r}^+|, r, {\bf x_1})S(|\phi_{t}^+\rangle\langle \phi_{t}^+|, t, {\bf x_2})\\
	+&\sum_{i=1}^{m}{p_i|c_i(r,t,+-)|^2 }\cdot S(|\phi_{r}^+\rangle \langle \phi_{r}^+|, r, {\bf x_1})S(|\phi_{t}^-\rangle\langle \phi_{t}^-|, t, {\bf x_2})\\
	+&\sum_{i=1}^{m}{p_i|c_i(r,t,-+)|^2 }\cdot S(|\phi_{r}^-\rangle \langle \phi_{r}^-|, r, {\bf x_1})S(|\phi_{t}^+\rangle\langle \phi_{t}^+|, t, {\bf x_2})\\
	+&\sum_{i=1}^{m}{p_i|c_i(r,t,--)|^2 }\cdot S(|\phi_{r}^-\rangle \langle \phi_{r}^-|, r, {\bf x_1})S(|\phi_{t}^-\rangle\langle \phi_{t}^-|, t, {\bf x_2})\\
	\end{split}
	\label{spectrum}
	\end{equation}

	In the standard Bell's inequality, the experimentally measured correlation is defined as:
	\begin{equation}
	E(r,t) = \frac{N(+,+)+N(-,-)-N(+,-)-N(-,+)}{N(+,+)+N(-,-)+N(+,-)+N(-,+)}
	\end{equation}
	where $N(+,+)$ represents number of measurements yielding '+' in both measurement r and measurement t of channel I and II.
	
	These measurements results are used to form the well-know Clauser-Horne-Shimony-Holt (CHSH) inequality\cite{clauser1969proposed}:
	\begin{equation}
	|E(r,t)+E(r,s)+E(q,s)-E(q,t)|\leq2
	\end{equation}
	
	For continuous measurement results, we construct an auxiliary function $V(r,t,{\bf x_1}, {\bf x_2})$ to simplify calculating the correlation function $E(r, t)$. The functional form of the auxiliary function could be found in the appendix.
	The generalized  Standard Bell's inequality for continuous variables takes the following form:
	\begin{equation}
	\begin{split}
	E =  &\left|\int{S(\rho, r, t, {\bf x_1}, {\bf x_2})V(r,t,{\bf x_1}, {\bf x_2})d{\bf x_1}d{\bf x_2}}+
	\int{S(\rho, q, t, {\bf x_1}, {\bf x_2})V(q,t,{\bf x_1}, {\bf x_2})d{\bf x_1}d{\bf x_2}}\right. \\
	+&\left.\int{S(\rho, r, s, {\bf x_1}, {\bf x_2})V(r,s,{\bf x_1}, {\bf x_2})d{\bf x_1}d{\bf x_2}}-
	\int{S(\rho, q, s, {\bf x_1}, {\bf x_2})V(q,s,{\bf x_1}, {\bf x_2})d{\bf x_1}d{\bf x_2}}\right|\leq 2
	\end{split} 
	\label{inequality}
	\end{equation}
	If this inequality, Eq.(\ref{inequality}), is violated, then the  system is entangled.

	\section{Experiment design and simulation results}
	
	Recently, Zare and coworkers reported the rotationally inelastic collisions between HD and H$_2$, D$_2$ molecules at very low temperature\cite{perreault2018cold}. In their scattering experiments, $|H\rangle$ (Orientation of molecules is parallel to its propagating direction) and $|V\rangle$ states (Orientation of molecules is vertical to its propagating direction) lead to very different scattering results, which offers us here a possible setup to classify entanglement between molecules based on the scattering results. In this section, we will propose an experimental setting to classify entanglement with continues measurement results.

	In the following simulations, we consider four measurements with respect to four  different sets of eigenstate basis. The first one is the scattering measurements corresponding to eigenstate $|H\rangle$ and $|V\rangle$, which we note as measurement $Z$.
	The second one is corresponding to the eigenstates $|+\rangle =\frac{1}{2} (|H\rangle +|V\rangle)$ and  $|-\rangle =\frac{1}{2} (|H\rangle -|V\rangle)$, which we note as measurement $X$.
	The other two measurements are taken with $\frac{Z+X}{\sqrt{2}}$ and $\frac{Z-X}{\sqrt{2}}$ basis.
	
	Measurement results in the $Z$-basis could be taken from Zare's experimental results, which are continuous as a function of the scattering angles\cite{perreault2018cold}. We assume that the measurement in the $X$-basis and $\frac{Z-X}{\sqrt{2}}$ are projection measurements, so that their results $\Gamma$ are discrete, which is very common in experimental set up. In order to check the derived inequality, Eq. (\ref{inequality}), we assume that there is another scattering experiment for measurement $\frac{Z+X}{\sqrt{2}}$, whose results satisfy the Gaussian distribution. In Fig.(1), we show the ideal results (Scattering angle $\theta$) of measurement $Z$ (upper part) and $\frac{Z+X}{\sqrt{2}}$ (lower part).
	
	\begin{figure}
		\begin{center}   
			\includegraphics[width=14cm,clip]{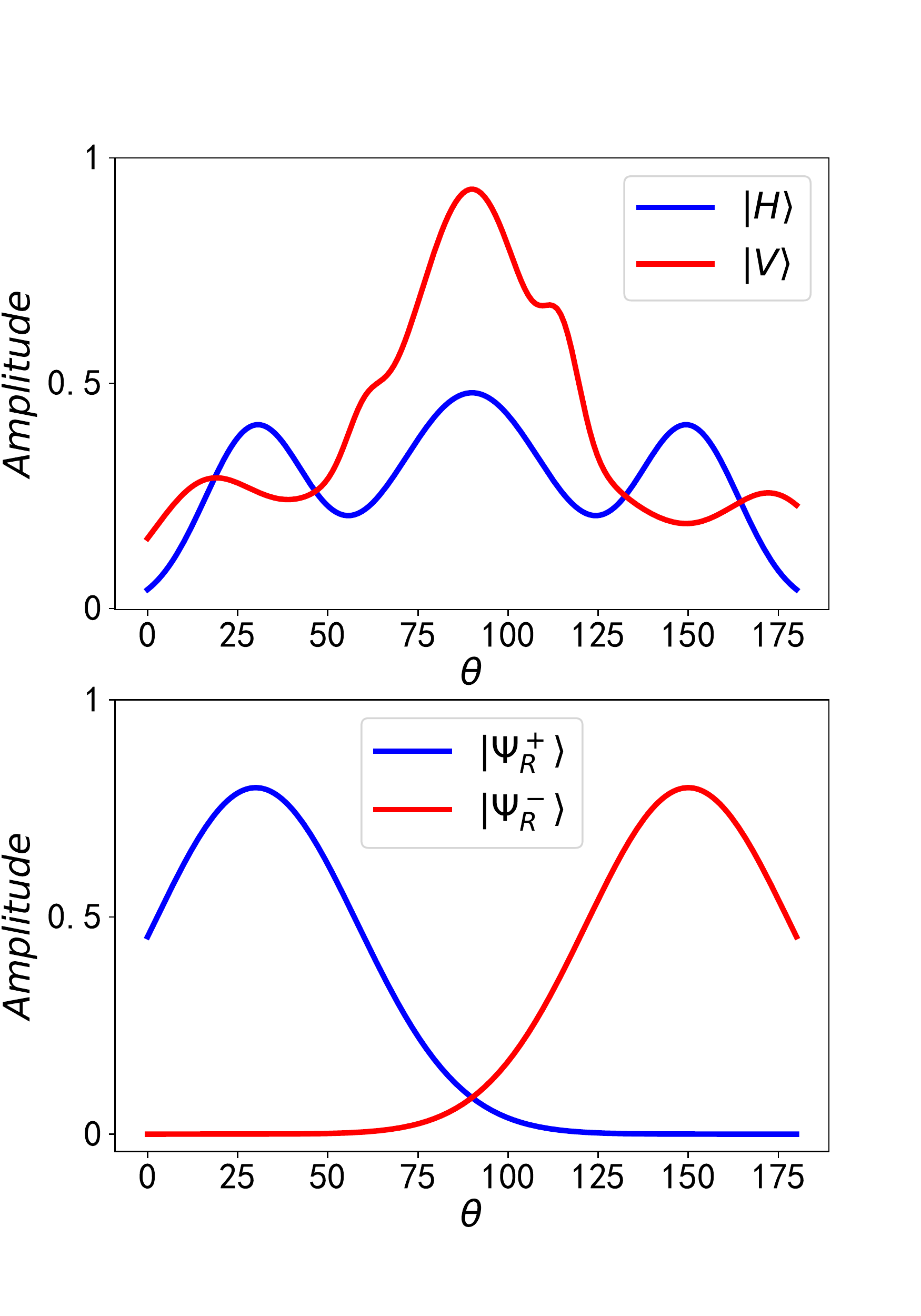}
			
			\caption{{\bf Ideal results of the measurement $Z$  (upper) and  measurement $\frac{Z+X}{\sqrt{2}}$  (lower)}\\	
				Results of measurement $Z$ (upper) are calculated from Ref\cite{perreault2018cold} Fig.(4a,b). The blue line $f_H(\theta)$ represents the scattering angle distribution of particles at state $|H\rangle$, and the red line $f_V(\theta)$ represents the scattering angle distribution of particles at state $|V\rangle$. For the measurement $\frac{Z+X}{\sqrt{2}}$, its two eigenstates are $|\Psi_R^+\rangle$ and $|\Psi_R^-\rangle$. We assume scattering angle distribution $f_\pm(\theta)$ of $|\Psi_R^+\rangle$ and $|\Psi_R^-\rangle$ satisfy Gaussian distribution. $f_+(\theta)\propto exp[-(\theta-30)^2/40^2]$\ (blue line in the lower figure), $f_-(\theta)\propto exp[-(\theta-150)^2/40^2]$\ (red line in the lower figure).
			}	
		\end{center} 
	\end{figure}

	For our simulations, we start by preparing the oriented HD molecules $|v=1, j=2, m=0\rangle$ in the state  $|H\rangle$, whose orientation is parallel to its propagation direction (y-axis in Fig.(2a)), and state $|V\rangle$, whose orientation is vertical (z-axis in Fig.(2a)). One molecule in group I and another  in group II are combined together, then prepared at different states (Werner state, superposition state, or mixed state), as shown in Fig.(2b).
	For each pair, HD molecules are divided for two channels. In channel I, they will scatter with H$_2$ clusters. If HD is collides with H$_2$, the orange sensors will measure the scattering particles, and the scattering results will be performed by measuremnet $Z$. If the HD does not collide with $H_2$ molecules, the gray sensor will carry on an $X$-measurement(Fig.(2c)). In channel II, the scattering process will go through measurement $\frac{Z+X}{\sqrt{2}}$, and HD that are not scattered are measured by $\frac{Z-X}{\sqrt{2}}$.
	
	\begin{figure}
		\begin{center}
			\includegraphics[width=14cm,clip]{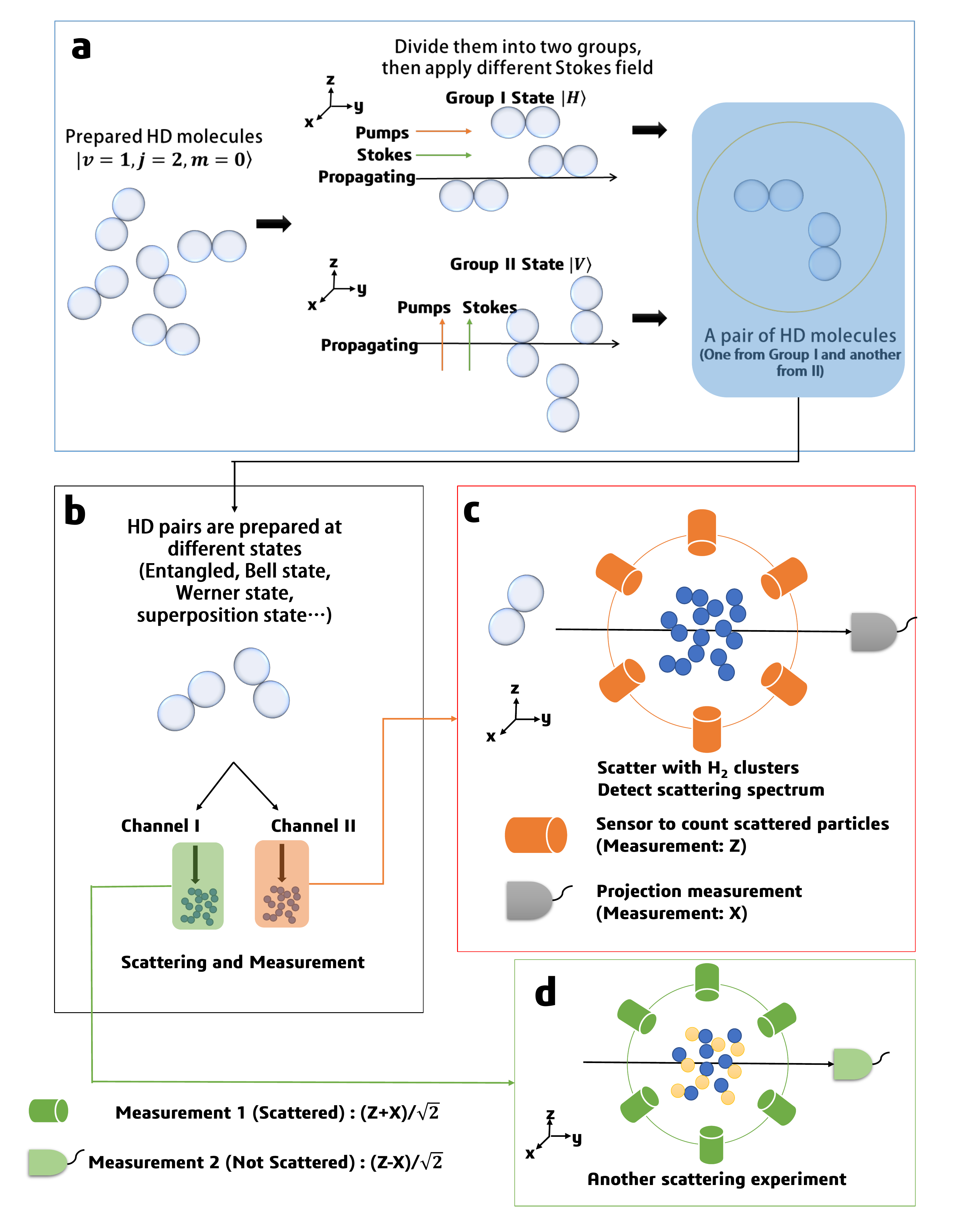}
			\caption{
				{\bf Sketch of the experiment design:}\\	
				HD molecules are prepared in the state $|v=1,j=2,m=0 \rangle$ using Stark-induced adiabatic Raman passage (SARP) \cite{perreault2018cold}. Then we can divide them into two molecular beams (two groups). If we apply the Pumps and Stokes electric field along y-axis, HD in group I are set at state $|H\rangle$ (Orientation of HD is along the y-axis, parallel to direction of its propagation). The molecule HD in group II are set at state $|V\rangle$ (Orientation of HD is along the z-axis, norm to direction of its propagation). One molecule from group I and another from group II combined together,  then HD pairs are prepared at different initial states (If we do nothing, these pairs will stay on a mixed state $|H\rangle \otimes |V\rangle$). The specific prepared state will go through two channels, molecule in channel I will scatter with H$_2$ clusters, where the bond axis of H$_2$ is distributed isotropically. Sensors (orange) are used to detect scattered particles and count numbers for each angle (Z-Measurement). Molecules that are not scattered will go to another sensor (gray), by which they will be measured on the eigenstates $|+\rangle$ and $|-\rangle$ (X-Measurement). In channel II, we set another experiment, so that scattered prepared HD molecules with isotropically distributed H$_2$ clusters are measured under $\frac{Z+X}{\sqrt{2}}$, while the others are measured under $\frac{Z-X}{\sqrt{2}}$.}	
		\end{center} 
	\end{figure}

	To perform the simulations, we are going to assume that one can prepare
	the states from the two oriented molecular beams into the combinations:
	$|H\rangle |H\rangle$;  $|V\rangle |V\rangle$; and  $|+\rangle|+\rangle $, $|-\rangle|-\rangle $ and finally, the Werner state:

	\begin{equation}
	\rho_w(p)=\frac{p}{2}(\ |HV \rangle+|VH \rangle\ )(\ \langle HV|+\langle  VH |\ )+\frac{1-p}{4}I
	\end{equation}
	
	The free parameter $p$ describes entanglement of Werner state. $p=0$ indicates separablelity,  entanglement for $p \leq 1/3$ and bell states of  maximum entanglement at $p=1$\cite{werner1989quantum}.

	If enough HD molecule pairs are prepared as mentioned above, the predicted spectra for different initial states are shown in Fig.(3). 
	The simulation is based on data of scattering experiment between HD and H$_2$ clusters taken from  Ref\cite{perreault2018cold}(Scattering spectrum Fig.(4a,b)).

	We can calculate the density matrix  $\rho_{w}$  for different measurements. For example, if both particles in channel I and II are scattered, the measurement in the basis $|H\rangle$ and $|V\rangle$ gives:
	
	\begin{equation}
	\begin{split}
	\rho_w=\begin{array}{@{}r@{}c@{}c@{}c@{}c@{}l@{}}
	& |HH\ \rangle & |HV\ \rangle & |VH\ \rangle & |VV\  \rangle  \\
	\left.\begin{array}
	{c} |HH\ \rangle \\ |HV\ \rangle \\ |VH\ \rangle \\ |VV\ \rangle \end{array}\right(
	& \begin{array}{c} \frac{1-p}{4} \\ 0 \\ 0\\ 0 \end{array}
	& \begin{array}{c} 0 \\ \frac{1+p}{4} \\ \frac{p}{2}\\ 0 \end{array}
	& \begin{array}{c} 0 \\ \frac{p}{2}\\ \frac{1+p}{4}\\ 0 \end{array}
	& \begin{array}{c} 0 \\ 0 \\ 0 \\ \frac{1-p}{4} \end{array}
	& \left)\begin{array}{c} \\ \\ \\  \\ \end{array}\right.
	\end{array}
	\end{split}
	\end{equation}
	
	If we change the basis set, then the density matrix in the basis $|+\rangle$ and $|-\rangle$ gives:
	
	\begin{equation}
	\begin{split}
	\rho_w=\begin{array}{@{}r@{}c@{}c@{}c@{}c@{}l@{}}
	& |++\ \rangle & |+-\ \rangle & |-+\ \rangle & |--\  \rangle  \\
	\left.\begin{array}
	{c} |++\ \rangle \\ |+-\ \rangle \\ |-+\ \rangle \\ |--\ \rangle \end{array}\right(
	& \begin{array}{c} \frac{1+p}{4} \\ 0 \\ 0\\ -\frac{p}{2} \end{array}
	& \begin{array}{c} 0 \\ \frac{1-p}{4} \\ 0\\ 0 \end{array}
	& \begin{array}{c} 0 \\ 0\\ \frac{1-p}{4}\\ 0 \end{array}
	& \begin{array}{c} -\frac{p}{2} \\ 0 \\ 0 \\ \frac{1+p}{4} \end{array}
	& \left)\begin{array}{c} \\ \\ \\  \\ \end{array}\right.
	\end{array}
	\end{split}
	\end{equation}
	We could do the same expansion for different measurements. 
	Based on Eq.(\ref{spectrum}),  we can predict their results of different initial states, as shown in Fig.(3).
	The first column represents the simulation results when molecules in both channels are scattered. The second column represents simulation results when HD goes through channel I is scattered but not for channel II. The third column represents simulation results when HD goes through channel II is scattered but not for channel I. And last column represents results when particle are not scattered.

	\begin{figure}
		\begin{center}
			\fbox{\includegraphics[width=15cm,clip]{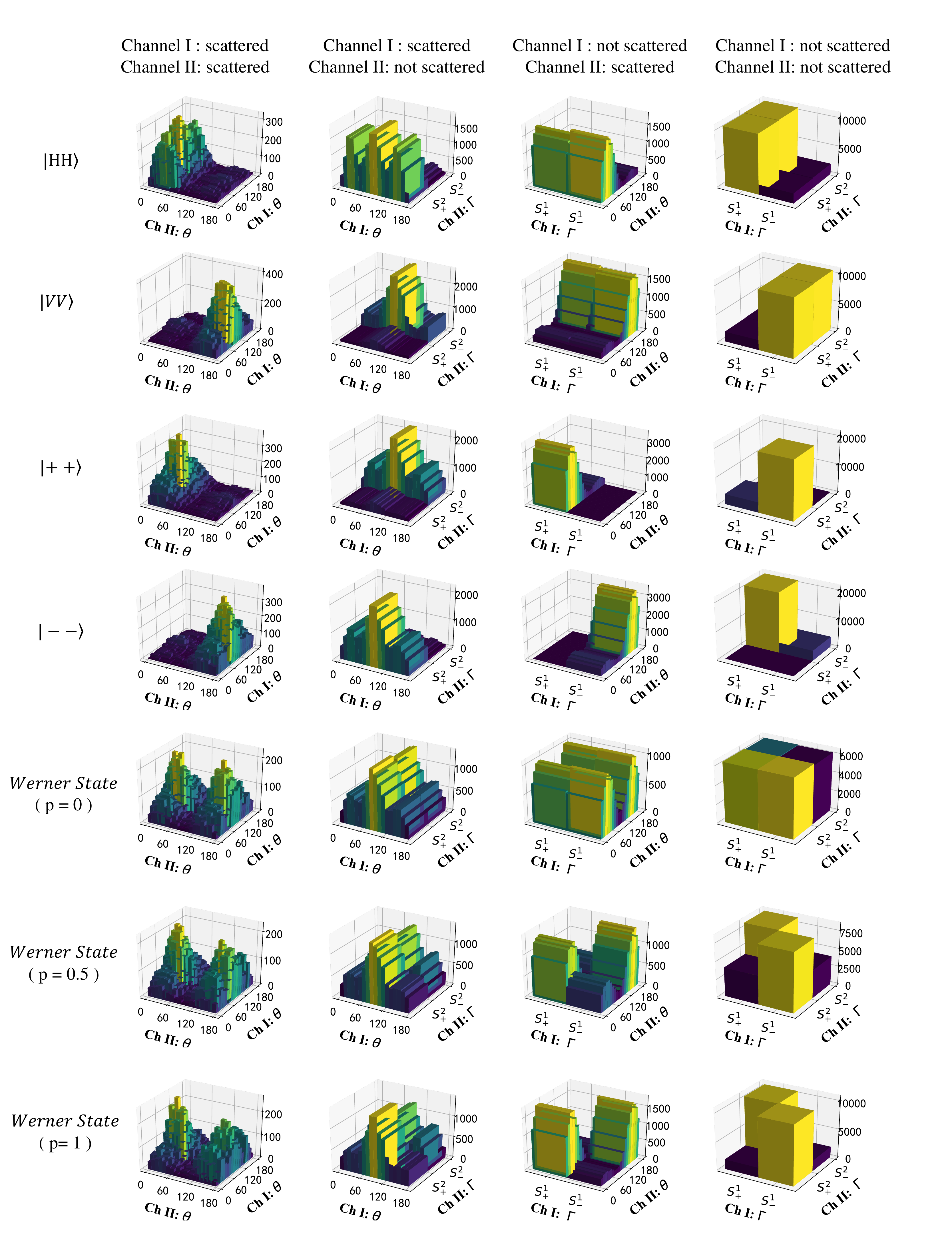}}
			
			\caption
			{
				{\bf Simulation results of the scattering experiments}\\	
				The figure shows histograms of the simulations count results (Z-axis in the Figure) as a function of different measures in  channel I and channel II.
				{\bf Row 1}: The spectrum for separable state $|HH\rangle$,  both HD molecules are in the state $|H\rangle$.
				{\bf Row 2}: The spectrum for separable state $|VV\rangle$.
				{\bf Row 3}: The spectrum for superposition state $|++\rangle$, where $|+\rangle= \frac{1}{\sqrt{2}}(|H\rangle+|D\rangle)$.
				{\bf Row 4}: The spectrum for superposition state $|--\rangle$, where $|-\rangle= \frac{1}{\sqrt{2}}(|H\rangle-|D\rangle)$.
				{\bf Row 5,6,7}: The spectrum for Werner state $\rho_w(p)$. When p=1, the prepared pairs are at Bell state $\frac{1}{\sqrt{2}}(|HD\rangle+|DH\rangle)$.}
		\end{center} 
	\end{figure}
	
	The state $|HH\rangle$ can be easily classified as its special spectrum in the first column (Ch I: scattered; Ch II: scatter), and State $|++\rangle$ shows significant difference with others in the second (Ch I: scattered; Ch II: not scatter) and third column (Ch I: not scattered; Ch II: scatter). For the Werner state, they share the same spectrum in the second and third column, yet their spectrum in the last column (Ch I: not scattered; Ch II: not scatter) offer us some feature: counts for $S^1_+S^2_+$ and $S^1_-S^2_-$ will decrease as $p$ increases. Also, the  difference in the first column, where there is a sub-peak (around $\theta_1 =30, \theta_2 = 90$) increases  when $p$ increases. Based on these features, it is possible to distinguish  them from each other. However, if the initial state is a complex mixed state, it is impossible for us to derive the initial state, because there are infinite possible results of  $\sum_{i=1}^{m}{p_i|c_i(r,t,\pm\pm)|^2 }$ only with restriction $\sum_{i=1}^{m}p_i=1 $ and $\sum_{j_{1,2}=\pm}|c_i(r,t,j_1, j_2)|^2=1 $. However, with more information of $p_i$, there is a chance to resolve the quantum state. For example,  if we have already know that the pairs are at Werner state, we could obtain the quantum state, as shown in the simulation work.
	
	\section{Simulation method}
	
	Here, we show how to obtain the simulation results. As an example we will take the first row in Fig.2 (State $|HH\rangle$, Channel I: scattered, Channel II: scattered). Consider the HD molecule that goes through Channel I. If the molecule HD scatted with a probability $P_{scatter}$, then  the Random Number Generator(RNG), where we assume that the RNG produces random numbers with uniform distribution,  produces a number $a_1$, $0\leq a_1\leq1$. If $a_1\leq P_{scatter}$, then this HD molecule with a specific state will collide with H$_2$ clusters, otherwise it will be measured by the gray sensor as shown in Fig.(2). For the scattered HD molecules, the scattering spectrum (measurement distribution) is $f_{H}(\theta)$, as shown in Fig.(1), where $\theta$ represents scattering angles, and we can get $f_{H}({\theta})$ by fitting the results in Zare's experimental measuremnets\cite{perreault2018cold}. If the HD molecule at state $|H\rangle$ is scattered, the following process is used to generate its scattering angle: A random angle $0\leq\theta\leq 180$ and a random number $a_2$, $0\leq a_2\leq1$ are produced be RNG, if $a_2\leq f_H(\theta)$, we accept $\theta$ as the scattering angle, otherwise this process is repeated. For the molecules that are not scattered, as $|H\rangle=\frac{1}{\sqrt{2}}(|+\rangle+|-\rangle)$, the measurement result has half possibility to be $S_+$ and another half to be $S_-$. We can use one random number to simulate the measurement result of these un-scattered molecules.
	
	\section{Discussion and conclusion}

	In the measurement setting of Werner state, Eq.(\ref{inequality}) is violated when $p > \frac{1}{\sqrt{2}}$. In the meanwhile, we know that Werner state will be entangled state when $p > \frac{1}{3}$. Hence, violation of Eq.(\ref{inequality}) guarantees existence of entanglement yet the non-violation of the inequality does not exclude the possibility of entanglement.

	\begin{figure}
		\begin{center}
			\includegraphics[width=15cm,clip]{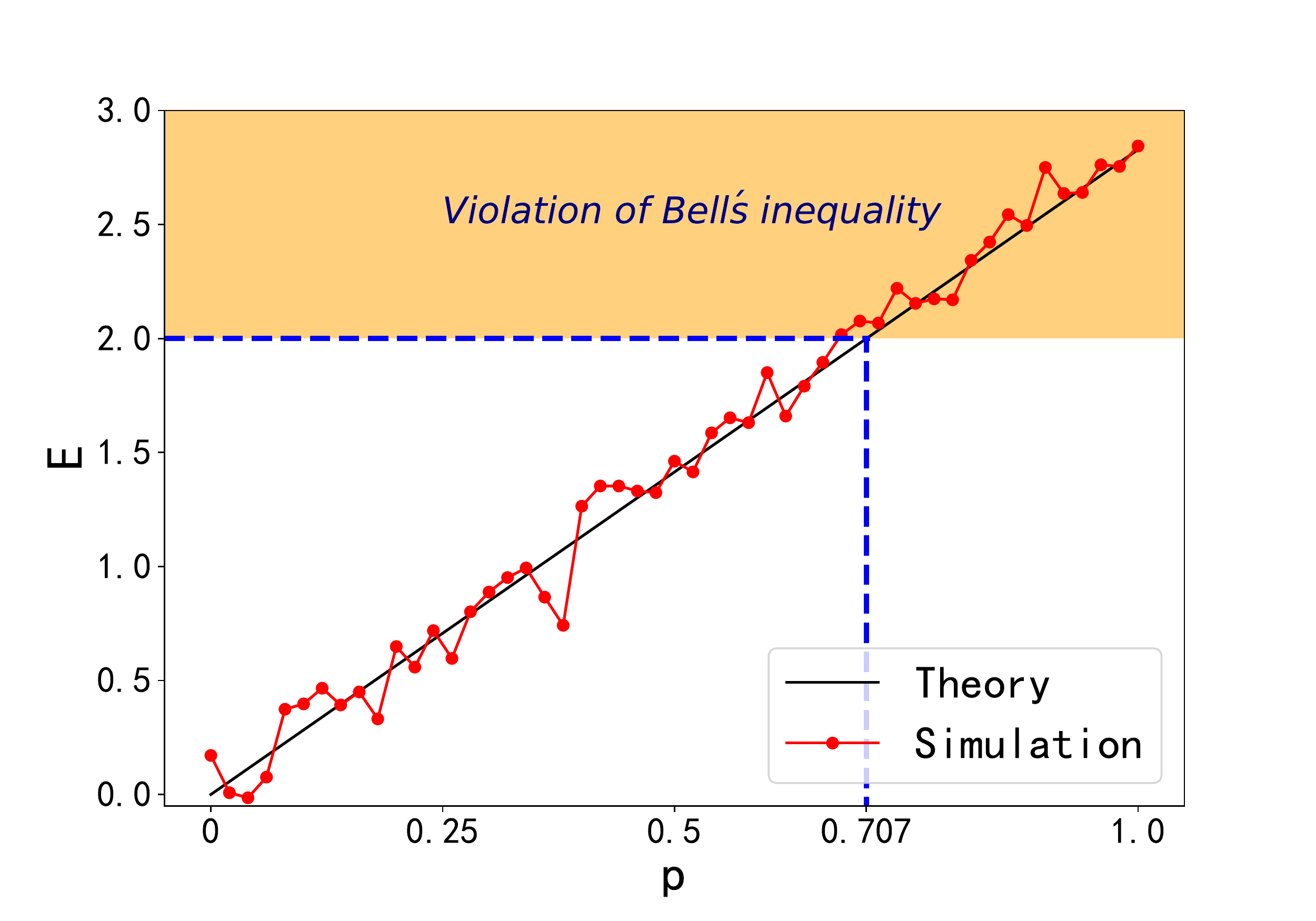}
			\caption{{\bf Theory and simulation results for $E$ of particle pairs at different Werner state $\rho_{w}(p)$}
				To calculate the integral, we divided scattering angles into 18 slots (10 degree per slot). We studied ~$1\times 10^6$ particles in simulation, and possibility to be scattered is set as 0.4. 
			}	
		\end{center} 
	\end{figure}
	As shown in Fig.(4), the simulation result (Red) $E$ is very close to the theory result (Black). The difference is mostly from the statistical error (Please refer to the appendix for method in simulation). We studied ~$1\times 10^6$ particles, and divided scattering angles into 18 slots uniformly. We also want to mention that if we divide the scattering angle into too many or too few slots ($\leq10$ or $\geq$50), simulation result $E$ will be much far away from the theory prediction.

	In summary, we have generalized the standard Bell’s inequality from discrete to continuous measurement results.  We designed an experiment setting as a potential application of violating the Bell’s inequality. We performed theoretical simulations to show the validity of the proposed experiment. 
	of Bell’s inequality.
	The method is general and might be used to design and classify entanglement in new molecular scattering experiments.

	\section*{Acknowledgments}
	The authors would like to thank Andrew Hu, Rongxin Xia and Teng Bian for useful discussion.

	{\bf Funding:} This material is based upon work supported by the U.S. Department of Energy, Office of Basic Energy Sciences, under Award
	Number DE-SC0019215. 
	{\bf Author contributions:} S.K. developed and supervised the project. J.L. performed the calculations.
	Both authors performed data analysis and modeling and contributed to the writing of the manuscript.
	{\bf Competing interests:} The authors declare that they no competing interests. Data and materials availability: All data needed to evaluate the conclusions in the paper are present in the paper and/or the Supplementary Materials. 
	Additional data related to this paper may be requested from the authors.
	
	\section*{Appendix}
	In this appendix we show how to build the auxiliary function $V(r,t,{\bf x_1}, {\bf x_2})$ to calculate the correlation function $E(r, t)$ from the continuous measurement results $r$ and $t$.
	For the continuous variables, we can define the correlation function as 
	\begin{equation}
	E(r,t) = \sum_{i=1}^{m}{p_i|c_i(r,t,++)|^2 } + \sum_{i=1}^{m}{p_i|c_i(r,t,--)|^2 } - \sum_{i=1}^{m}{p_i|c_i(r,t,+-)|^2 } - \sum_{i=1}^{m}{p_i|c_i(r,t,-+)|^2 }
	\end{equation}
	
	Generally, $\{S(|\phi_{r}^{j_1}\rangle\langle\phi_{r}^{j_1}|, r, {\bf x_1}) S(|\phi_{t}^{j_2}\rangle\langle\phi_{t}^{j_2}|, t, {\bf x_2})\}$ are linearly independent, which makes it possible to calculate $\sum_{i=1}^{m}{p_i|c_i(r,t,j_1,j_2)|^2}, (j_1,j_2=+,-)$.
	For arbitrary measurement $r$, we have
	\begin{equation*}
	\int{S(|\phi_{r}^\pm\rangle\langle \phi_{r}^\pm|, r, {\bf x})d{\bf x}} = 1
	\end{equation*}
	
	To simplify the analysis, we defined the following functions: 	
	\begin{equation}
	v(r,+,{\bf x_1}) = \frac{S(|\phi_{r}^+\rangle\langle \phi_{r}^+|, r, {\bf x_1})
		-\int{S(|\phi_{r}^+\rangle\langle \phi_{r}^+|, r, {\bf x_1})S(|\phi_{r}^-\rangle\langle \phi_{r}^-|, r, {\bf x_1})d{\bf x_1}}}
	{\int{S(|\phi_{r}^+\rangle\langle \phi_{r}^+|, r, {\bf x_1})S(|\phi_{r}^+\rangle\langle \phi_{r}^+|, r, {\bf x_1})d{\bf x_1}}
		- \int{S(|\phi_{r}^+\rangle\langle \phi_{r}^+|, r, {\bf x_1})S(|\phi_{r}^-\rangle\langle \phi_{r}^-|, r, {\bf x_1})d{\bf x_1}}}
	\end{equation}
	\begin{equation}
	v(r,-,{\bf x_1}) = \frac{S(|\phi_{r}^-\rangle\langle \phi_{r}^-|, r, {\bf x_1})
		-\int{S(|\phi_{r}^+\rangle\langle \phi_{r}^+|,r, {\bf x_1})S(|\phi_{r}^-\rangle\langle \phi_{r}^-|, r, {\bf x_1})d{\bf x_1}}}
	{\int{S(|\phi_{r}^-\rangle\langle \phi_{r}^-|, r, {\bf x_1})S(|\phi_{r}^-\rangle\langle \phi_{r}^-|, r, {\bf x_1})d{\bf x_1}}
		- \int{S(|\phi_{r}^+\rangle\langle \phi_{r}^+|, r, {\bf x_1})S(|\phi_{r}^-\rangle\langle \phi_{r}^-|, r, {\bf x_1})d{\bf x_1}}}
	\end{equation}
	
	\begin{equation}
	v(t,+,{\bf x_2}) = \frac{S(|\phi_{t}^+\rangle\langle \phi_{t}^+|, t, {\bf x_2})
		-\int{S(|\phi_{t}^+\rangle\langle \phi_{t}^+|, t, {\bf x_2})S(|\phi_{t}^-\rangle\langle \phi_{t}^-|, t, {\bf x_2})d{\bf x_2}}}
	{\int{S(|\phi_{t}^+\rangle\langle \phi_{t}^+|, t, {\bf x_2})S(|\phi_{t}^+\rangle\langle \phi_{t}^+|, t, {\bf x_2})d{\bf x_2}}
		- \int{S(|\phi_{t}^+\rangle\langle \phi_{t}^+|, r, {\bf x_2})S(|\phi_{t}^-\rangle\langle \phi_{t}^-|, r, {\bf x_2})d{\bf x_2}}}
	\end{equation}
	\begin{equation}
	v(t,-,{\bf x_2}) = \frac{S(|\phi_{t}^-\rangle\langle \phi_{t}^-|, t, {\bf x_2})
		-\int{S(|\phi_{t}^+\rangle\langle \phi_{t}^+|,t, {\bf x_2})S(|\phi_{t}^-\rangle\langle \phi_{t}^-|, t, {\bf x_2})d{\bf x_2}}}
	{\int{S(|\phi_{t}^-\rangle\langle \phi_{t}^-|, t, {\bf x_2})S(|\phi_{t}^-\rangle\langle \phi_{t}^-|, t, {\bf x_2})d{\bf x_2}}
		- \int{S(|\phi_{t}^+\rangle\langle \phi_{t}^+|, t, {\bf x_2})S(|\phi_{t}^-\rangle\langle \phi_{t}^-|, t, {\bf x_2})d{\bf x_2}}}
	\end{equation}
	
	with 
	\begin{equation}
	\int{v(r,\pm,{\bf x_1})S(|\phi_{r}^\pm\rangle\langle \phi_{r}^\pm|, r, {\bf x_1})d{\bf x_1}} = 1
	\end{equation}
	and
	\begin{equation}
	\int{v(r,\pm,{\bf x_1})S(|\phi_{r}^\mp\rangle\langle \phi_{r}^\mp|, r, {\bf x_1})d{\bf x_1}} = 0
	\end{equation}
	Then, the correlation function is given by:
	\begin{equation}
	E(r,t) = \int {S(\rho, r, t, {\bf x_1}, {\bf x_2}) {\left[ v(r,+,{\bf x_1})-v(r,-,{\bf x_1}) \right]}
		{\left[v(t,+,{\bf x_2})-v(t,-,{\bf x_2})\right]} d{\bf x_1}d{\bf x_2}}
	\label{Eofinequality}
	\end{equation}
	Note that $\rho$ in Eq.(\ref{Eofinequality}) represents density matrix of bipartite systems.
	We define the auxiliary function as  
	\begin{equation}
	V(r,t,{\bf x_1}, {\bf x_2}) = {\left[ v(r,+,{\bf x_1})-v(r,-,{\bf x_1}) \right]}
	{\left[v(t,+,{\bf x_2})-v(t,-,{\bf x_2})\right]}
	\end{equation}
	\bibliographystyle{plain}
	\bibliography{ref}
	\newpage
	
	\section*{Supplementary Material}
	
	\subsection*{A. Details of the simulation method:}
	
	For simplicity, the density matrix of the two particles discussed in the text, Eq. (\ref{densityMatrix}) and Eq. (\ref{state}) can be written as,
	\begin{equation}
	\rho_{r,t}=\left[
	\begin{matrix}
	& A_{rt}^{++} & B_{rt}^{1,2} & B_{rt}^{1,3} & B_{rt}^{1,4}\quad \\
	& B_{rt}^{2,1} & A_{rt}^{+-} & B_{rt}^{2,3} & B_{rt}^{2,4}\quad \\
	& B_{rt}^{3,1} & B_{rt}^{3,2} & A_{rt}^{-+} & B_{rt}^{3,4}\quad \\
	& B_{rt}^{4,1} & B_{rt}^{4,2} & B_{rt}^{4,3} & A_{rt}^{--}\quad \\
	\end{matrix}
	\right]
	\label{densityM2}
	\end{equation}
	where the diagonal elements $A$  are real and the off diagonal elements $B$  are complex:
	\begin{equation}
	A_{rt}^{j_1,j_2}=\sum_{i=1}^{m}p_i|c_i(r,t,j_1j_2)|^2 \qquad j_i, j_2=+,-
	\end{equation}
	\begin{equation}
	\begin{split}
	B_{rt}^{n_1,n_2}=\sum_{i=1}^{m}p_ic_i(r,t,J(n_1))c_i(r,t,J(n_2))^*\\
	where \quad n_{1},n_2=1,2,3,4,\ n_1\neq n_2\\
	and \quad J(1)=++,\ J(2)=+-,\ J(3)=-+,\ J(4)=--
	\end{split}
	\end{equation}
	with  $B_{rt}^{n_1,n_2}={B_{rt}^{n_2,n_1}}^*$.
	All $A$ components in Eq.(\ref{densityM2}) can be obtained by analyzing the spectrum $S(\rho, r, t, {\bf x_1}, {\bf x_2})$. One possible method is to find the set $A_{rt}^{j_1,j_2}$, with which the function $D({\bf A_{rt}})$ would be minimum, where
	\begin{equation}
	D({\bf A_{rt}})=\int d{\bf x_1}\int d{\bf x_2}{\left[S(\rho, r, t, {\bf x_1}, {\bf x_2})-\sum_{j_1,j_2}A_{rt}^{j_1j_2}S(|\phi_{r}^{j_1}\rangle, r, {\bf x_1})S(|\phi_{t}^{j_2}\rangle, t, {\bf x_2})\right]}^2
	\end{equation} 
	
	To obtain the off-diagonal components in the density matrix, other measurements are needed. For particles going through channel I, the $r-$ and $q-$measurements are used, while $s-$ and $t-$measurement are used for particles from channel II. The relationship between eigenstates of the two different measurements can be written as:
	\begin{equation}
	\left[
	\begin{matrix}
	& |\phi_{k_1}^+\rangle\quad \\
	& |\phi_{k_1}^+\rangle\quad 
	\end{matrix}
	\right]
	=
	\left[
	\begin{matrix}
	& a_{k_1,k_2}^{++} & a_{k_1,k_2}^{+-}\quad \\
	& a_{k_1,k_2}^{-+} & a_{k_1,k_2}^{--}\quad 
	\end{matrix}
	\right]
	\left[
	\begin{matrix}
	& |\phi_{k_2}^+\rangle\quad \\
	& |\phi_{k_2}^+\rangle\quad 
	\end{matrix}
	\right]
	\end{equation}
	For simplicity, we can write
	\begin{equation*}
	{\bf C_i}(r,t)=
	\left[
	\begin{matrix}
	& c_i(r,t,++) &c_i(r,t,+-)\quad \\
	& c_i(r,t,-+) &c_i(r,t,--)\quad
	\end{matrix}
	\right]
	\end{equation*}
	If  we choose a different measurement in one channel, the matrix ${\bf C_i}(r,t)$ becomes
	\begin{equation}
	{\bf C_i}(r,s)={\bf C_i}(r,t)\cdot
	\left[
	\begin{matrix}
	& a_{t,s}^{++} & a_{t,s}^{+-}\quad \\
	& a_{t,s}^{-+} & a_{t,s}^{--}\quad 
	\end{matrix}
	\right]
	\label{relation}
	\end{equation}
	and similarly, 
	\begin{equation}
	{\bf C_i}(q,t)=
	\left[
	\begin{matrix}
	& a_{r,q}^{++} & a_{r,q}^{+-}\quad \\
	& a_{r,q}^{-+} & a_{r,q}^{--}\quad 
	\end{matrix}
	\right]\cdot{\bf C_i}(r,t).
	\end{equation}
	When $a_{st}^{+-}+a_{st}^{-+}=0$, some off-diagonal components in density matrix can be calculated:
	\begin{equation}
	\begin{split}
	B_{rt}^{1,2}&=\sum_{i=1}^{m}p_ic_i(r,t,++)c_i(r,t,--)^*\\
	&=\sum_{i=1}^{m}p_i|c_i(r,s,++)|^2\cdot|a_{st}^{++}|^2+\sum_{i=1}^{m}p_i|c_i(r,s,--)|^2\cdot|a_{st}^{--}|^2
	\end{split}
	\end{equation}
	and by the same method one can calculate $B_{rt}^{1,3},B_{rt}^{2,4},B_{rt}^{3,4}$.
	
	From Eq.(\ref{densityMatrix}) and Eq.(\ref{relation}), we can write the connection between the descriptions of the density matrix with different eigenstate basis (under different measurement) as:
	\begin{equation}
	\rho_{q,s}=
	\left[
	\begin{matrix}
	& a_{r,q}^{++} & a_{r,q}^{+-}\quad \\
	& a_{r,q}^{-+} & a_{r,q}^{--}\quad 
	\end{matrix}
	\right]\otimes
	\left[
	\begin{matrix}
	& a_{t,s}^{++} & a_{t,s}^{+-}\quad \\
	& a_{t,s}^{-+} & a_{t,s}^{--}\quad 
	\end{matrix}
	\right]
	\cdot\rho_{r,t}\cdot
	\left[
	\begin{matrix}
	& a_{r,q}^{++} & a_{r,q}^{+-}\quad \\
	& a_{r,q}^{-+} & a_{r,q}^{--}\quad 
	\end{matrix}
	\right]^\dagger\otimes
	\left[
	\begin{matrix}
	& a_{t,s}^{++} & a_{t,s}^{+-}\quad \\
	& a_{t,s}^{-+} & a_{t,s}^{--}\quad 
	\end{matrix}
	\right]^\dagger
	\label{relationF}
	\end{equation}

	If we could know every component of the density matrix, there would be more choice to classify entanglement.
	According to the PPT criterion\cite{peres1996separability}\cite{horodecki2001separability},  if $\rho^{T_B}$ has any negative eigenvalue,  then these particles are entangled.

	\subsection*{B. An example of spin $\frac{1}{2}$ particles:}

	Here, we will give an example to show how the simulation method works for spin $\frac{1}{2}$ particles. Although the simulation method  designed mainly for continuous measurement results, it will also show a good performance for discrete measurement results. 
	
	When spin $\frac{1}{2}$ particles  go through Stern-Gerlach (SG) apparatus whose magnetic field is along the z-axis, they will split into two beams along z-axis corresponding to eigenstate $|\uparrow\ \rangle$ and $|\downarrow\ \rangle$. When these particles go through SG apparatus with magnetic field along the x-axis, they will split into two beams along the x-axis corresponding to eigenstate $|+\rangle$ and $|-\rangle$.
	Assume that there is a machine that could produce particle pairs at some certain quantum states propagating along the y-axis. Each pair are separated and sent to two channels as shown in Fig.(5). Particles would then go through SG apparatus with magnetic field along x or z-axis randomly, by which four spectrum measurements could be collected.
	
	\begin{figure}
		\begin{center}
			\fbox{\includegraphics[width=14cm,clip]{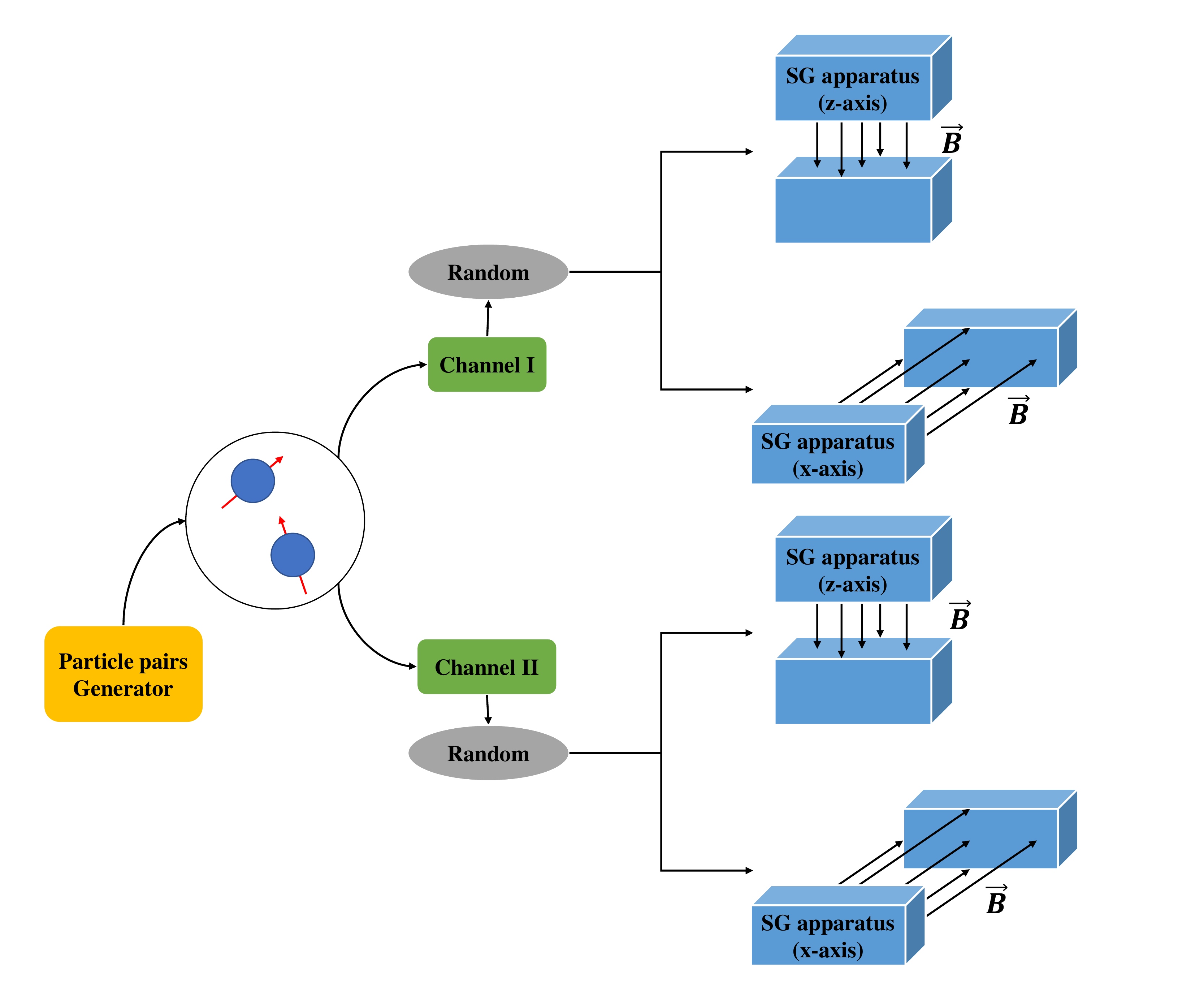}}
			\caption{{\bf Sketch of Bell's experiment of spin-$\frac{1}{2}$ particles}: The prepared spin-$\frac{1}{2}$ particles are divided into two channels, in each one they will go through SG-apparatus with magnetic field along the z-axis or the x-axis randomly. Then the count numbers of particle pairs are related to different measurement results. }	
		\end{center} 
	\end{figure}
	
	At the initial stage, we can prepare the spin-$\frac{1}{2}$ particle pairs in superposition, entangled, or mixed states as follow:
	
	{\bf Case I Superposition state:}
	Particle pairs produced in equal superposition
	\begin{equation}
	|\Psi_{s}\rangle=\frac{1}{2} [\ |\uparrow\uparrow\ \rangle + |\uparrow\downarrow\ \rangle + |\downarrow\uparrow\ \rangle + |\downarrow\downarrow\ \rangle\ ]
	\end{equation}

	{\bf Case II Entangled state:}
	Particle pairs are all at Bell state 
	\begin{equation}
	|\Psi_{e}\rangle=\frac{1}{\sqrt{2}}(|\uparrow\downarrow\ \rangle+|\downarrow\uparrow\ \rangle)
	\end{equation}
	
	{\bf Case III Mixed state:}
	Every pair produced consists of  one particle at state $|\uparrow\rangle$, and another one at state $|\uparrow\rangle$. These particles could be  describe by  using the density matrix: 
	\begin{equation}
	\rho_m=\frac{1}{2}\left[\ |\uparrow\downarrow\ \rangle\langle\uparrow\downarrow |+|\downarrow\uparrow\ \rangle\langle\downarrow\uparrow |\ 
	\right]
	\end{equation}
	After certain measurements, the spectrum for these cases could be simulated  based on Eq.(\ref{spectrum}).
	For the case I, the superposition state could be transformed to a different basis:
	\begin{equation}
	\begin{split}
	|\Psi_{s}\rangle&= \frac{1}{2} [\ |\uparrow\uparrow\ \rangle + |\uparrow\downarrow\ \rangle + |\downarrow\uparrow\rangle + |\downarrow\downarrow\ \rangle\ ]\\
	&= \frac{1}{\sqrt{2} } [\ |\uparrow+\rangle + |\downarrow+\rangle\ ]\\
	&= \frac{1}{\sqrt{2} } [\ |+\uparrow\rangle + |+\downarrow\ \rangle\ ]\\
	&= |++\rangle\\
	\end{split}
	\end{equation}
	Then it is convenient to calculate its spectrum under different measurement. If both of the SG apparatus in channel I and II have magnetic field along z-axis, then we obtain:
	\begin{eqnarray}
	\begin{split}
	S(|\Psi_{s}\rangle\langle\Psi_{s}|, z, z, {\bf x_1}, {\bf x_2})= 
	\ &\frac{1}{4} S(|\uparrow\rangle\langle\uparrow\ |, z, {\bf x_1})S(|\uparrow\ \rangle\langle \uparrow\ |, z, {\bf x_2}) \\
	+&\frac{1}{4}S(|\uparrow\ \rangle\langle \uparrow\ |, z, {\bf x_1})S(|\downarrow\ \rangle\langle \downarrow\ |, z, {\bf x_2}) \\
	+&\frac{1}{4}S(|\downarrow\ \rangle\langle \downarrow\ |, z, {\bf x_1})S(|\uparrow\ \rangle\langle \uparrow\ |, z, {\bf x_2}) \\
	+&\frac{1}{4}S(|\downarrow\ \rangle\langle \downarrow\ |, z, {\bf x_1})S(|\downarrow\ \rangle\langle \downarrow\ |, z, {\bf x_2})
	\end{split}
	\label{transform}
	\end{eqnarray}
	Eq.(\ref{transform}) explains the nearly same results in the spectrum measurements of $|\Psi_{s}\rangle$ in Fig.(7). And from Eq.(\ref{spectrum}), we could calculate its spectrum under other measurements as:
	\begin{eqnarray}
	\begin{split}
	S(|\Psi_{s}\rangle\langle\Psi_{s}|, z, x, {\bf x_1}, {\bf x_2})= 
	\ &\frac{1}{2} S(|\uparrow\ \rangle\langle \uparrow\ |, z, {\bf x_1})S(|+\rangle\langle +|, x, {\bf x_2}) \\
	+&\frac{1}{2}S(|\uparrow\ \rangle\langle \uparrow\ |, z, {\bf x_1})S(|-\rangle\langle -|, x, {\bf 	x_2}) 
	\end{split}\\
	\begin{split}
	S(|\Psi_{s}\rangle\langle\Psi_{s}|, x, z, {\bf x_1}, {\bf x_2})= 
	\ &\frac{1}{2} S(|+\rangle\langle +|, x, {\bf x_1})S(|\uparrow\ \rangle\langle \uparrow\ |, z, {\bf x_2}) \\
	+&\frac{1}{2}S(|-\rangle\langle -|, x, {\bf x_1})S(|\uparrow\ \rangle\langle \uparrow\ |, z, {\bf x_2}) 
	\end{split}
	\end{eqnarray}
	These equations explain why there are only two columns (histogram) left in spectrum of $|\Psi_{s}\rangle$ in Fig.(7) and Fig.(8). Since that $|\Psi_{s}\rangle=|++\rangle$ is already an eigenstate under the measurement $1x2x$ in which both of the magnetic field in channel I and II are along x-axis, we can directly get its spectrum $S(|\Psi_{s}\rangle\langle\Psi_{s}|, x, x, {\bf x_1}, {\bf x_2})= 
	S(|+\rangle\langle +|, x, {\bf x_1})S(|+\rangle\langle +|, x, {\bf x_2})$, as shown in $Fig.(6)$.
	
	Similarly, we calculated the spectrum of case 2 and 3, whose spectrum under different measurements are shown in Fig.(6-9) together with spectrum of $|\Psi_{s}\rangle$.
	\begin{figure}
		\begin{center}
			\fbox{\includegraphics[width=14cm,clip]{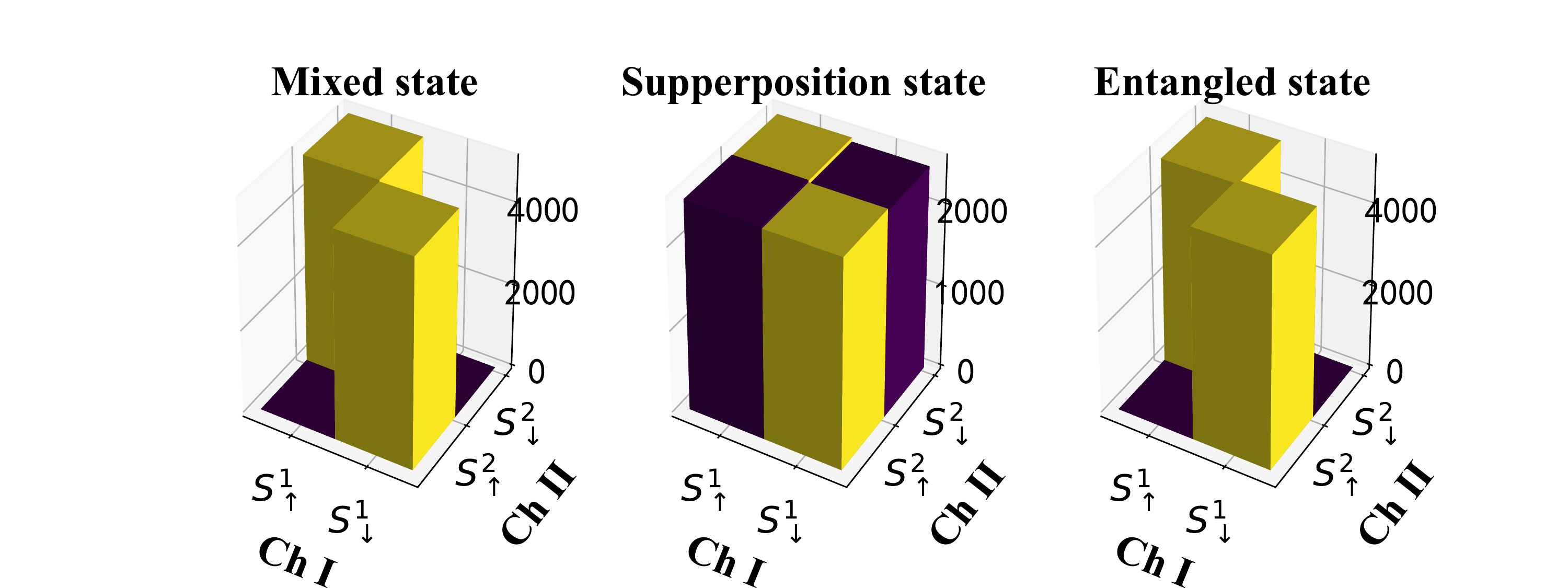}}
			
			\caption{{\bf Simulation result (1z2z)}\\
				We simulate the count number along Channel I and Channel II under the same measurement, where the magnetic field in SG apparatus of channel I and II are both along the z-axis (1z2z). $S^{1,2}_{\uparrow,\downarrow}$ represent particles that show state $|\uparrow\ \rangle$ or $\downarrow\ \rangle$ in channel I or channel II.
				Three figures represents simulation results of the three cases mentioned (Left: Mixed state; Middle: Superposition state; Right: Entangled state).}	
		\end{center} 
	\end{figure}
	
	\begin{figure}
		\begin{center}
			\fbox{\includegraphics[width=14cm,clip]{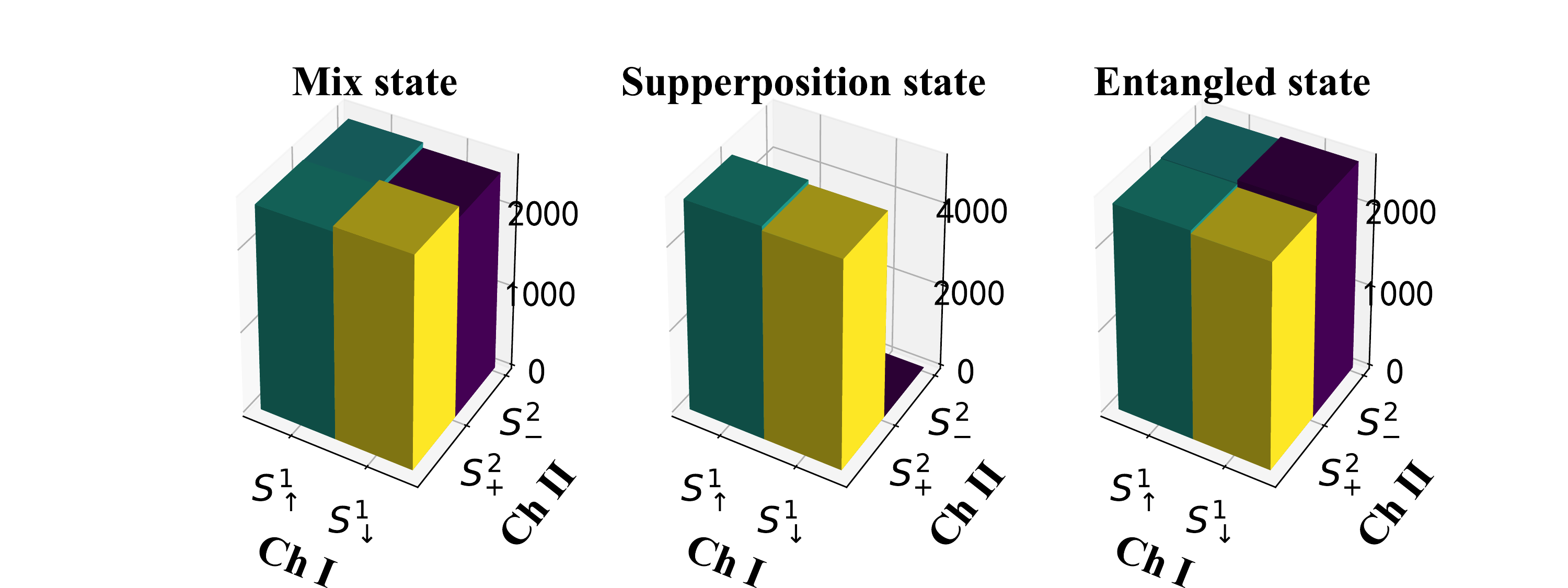}}
			\caption{{\bf Simulation result (1z2x)}\\
				We simulate the count number along Channel I and Channel II under the same measurement, where the magnetic field in SG apparatus of channel I is along z-axis, and II is along x-axis (1z2x). 
				$S^{1}_{\uparrow,\downarrow}$ represent the particles that show state $|\uparrow\ \rangle$ or $|\downarrow\ \rangle$ in channel I, and $S^{2}_{\pm}$ represent the particles that show state $|+\rangle$ or $|-\rangle$ in channel I.
				Three figures represents simulation results of the three cases mentioned (Left: Mixed state; Middle: Superposition state; Right: Entangled state).}	
		\end{center} 
	\end{figure}
	
	\begin{figure}
		\begin{center}
			\fbox{\includegraphics[width=14cm,clip]{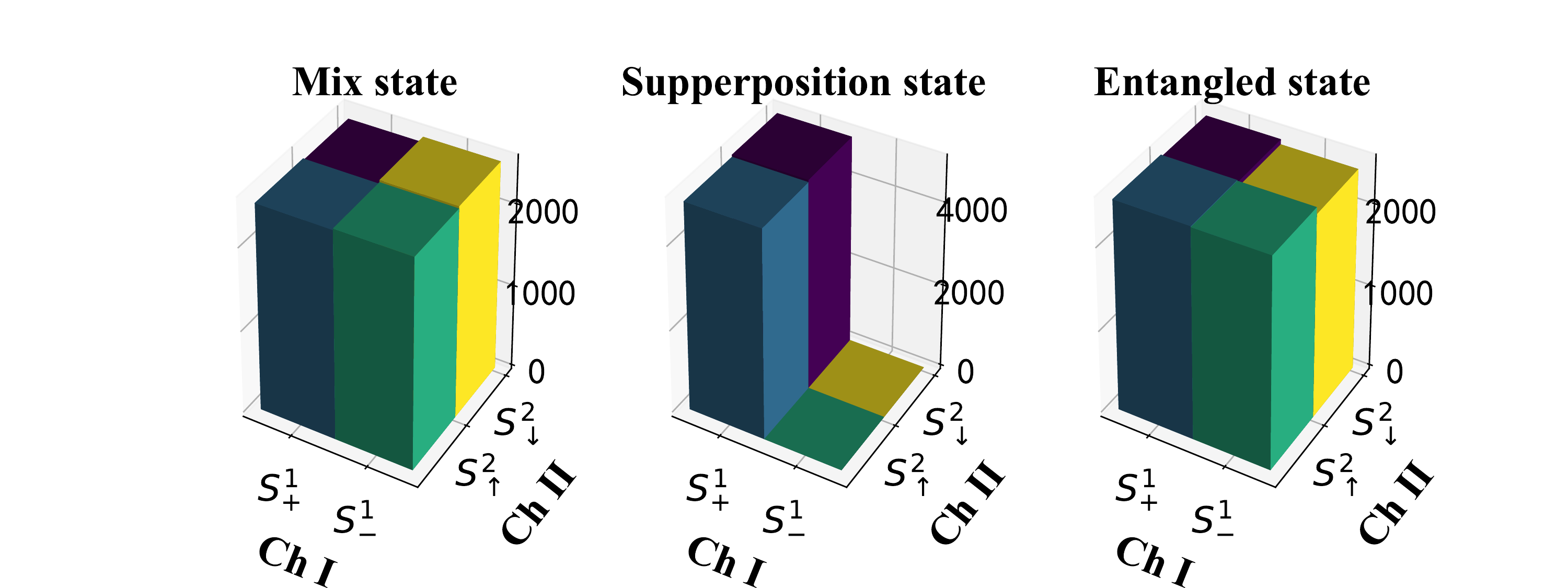}}
			\caption{{\bf Simulation result (1x2z)}\\
				We simulate the count number along Channel I and Channel II under the same measurement, where the magnetic field in SG apparatus of channel I is along x-axis, and II is along z-axis (1x2z). 
				$S^{1}_{\pm}$ represent the particles that show state $|+\rangle$ or $|-\rangle$ in channel I, and $S^{2}_{\uparrow,\downarrow}$ represent the particles that show state $|\uparrow\ \rangle$ or $|\downarrow\ \rangle$ in channel II.
				Three figures represents simulation results of the three cases mentioned (Left: Mixed state; Middle: Superposition state; Right: Entangled state).}	
		\end{center} 
	\end{figure}
	
	\begin{figure}
		\begin{center}
			\fbox{\includegraphics[width=14cm,clip]{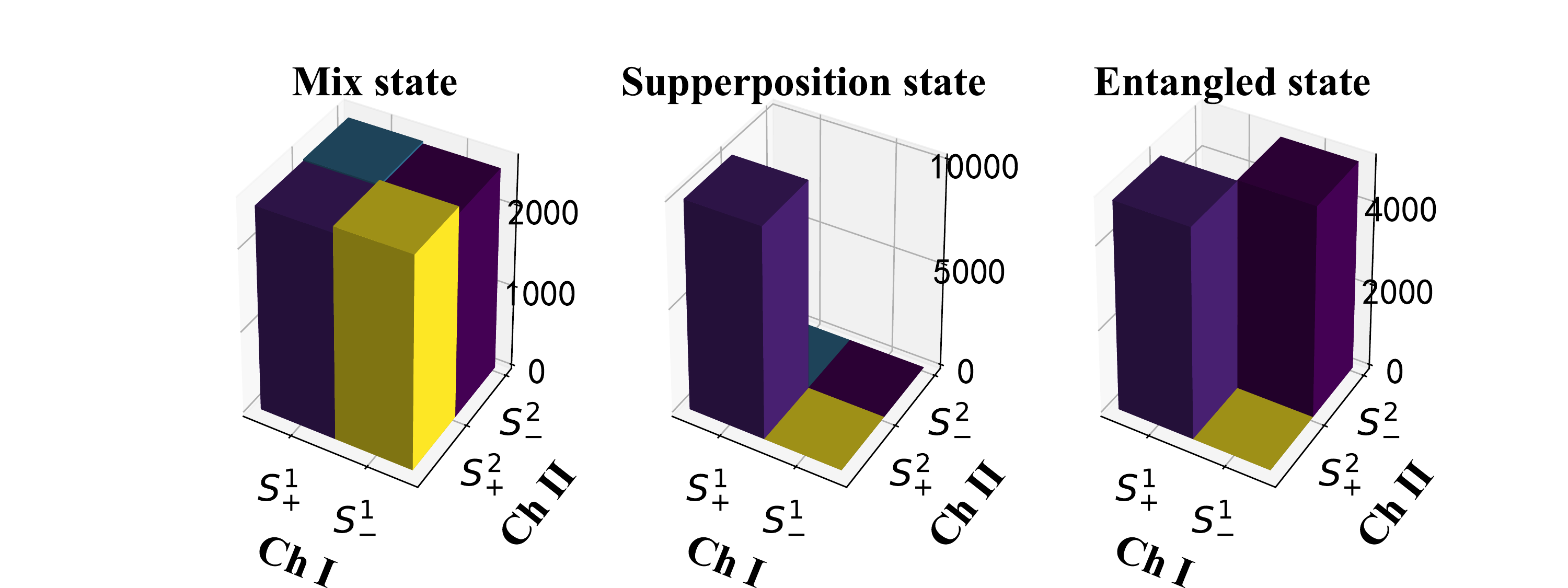}}
			\caption{{\bf Simulation result (1x2x)}\\
				We simulate the count number along Channel I and Channel II under the same measurement, where the both the magnetic field in SG apparatus of channel I and II are along x-{\tiny }axis (1x2x).
				$S^{1,2}_{\pm}$ represent the particles that show state $|+\rangle$ or $|-\rangle$ in channel I or channel II.
				Three figures represents simulation results of the three cases mentioned (Left: Mixed state; Middle: Superposition state; Right: Entangled state).}	
		\end{center} 
	\end{figure}
	
	As we can see from the simulation results, we can not distinguish these three states only from the spectrum $S(z,z,{\bf x_1},{\bf x_2})$, yet when we take all the four spectrum measurements into account, it is possible to classify them.

	Next we will discuss measurements starting from Warner state, which plays an important role in quantum teleportation\cite{lee2000entanglement}\cite{yeo2002teleportation}. Assume that a machine $M$ could generate particles at Werner state of the form:
	\begin{equation}
	\rho_w(p)=\frac{p}{2}(\ |\uparrow\downarrow\ \rangle+|\downarrow\uparrow\ \rangle\ )(\ \langle\  \downarrow \uparrow|+\langle\  \uparrow \downarrow |\ )+\frac{1-p}{4}I
	\end{equation}
	According to PPT-criterion, when $p < \frac{1}{3}$, particles are in separated states; while when $p\geq\frac{1}{3}$ they are entangled. 
	One can write the density matrix of such state in different basis according to various measurements. In the basis of $1z2z$ (both magnetic fields in channel I and II are along z-axis), we can rewrite the density matrix as
	\begin{equation}
	\begin{split}
	\rho=\begin{array}{@{}r@{}c@{}c@{}c@{}c@{}l@{}}
	& |\uparrow\uparrow\ \rangle & |\uparrow\downarrow\ \rangle & |\downarrow\uparrow\ \rangle & |\downarrow\downarrow\  \rangle  \\
	\left.\begin{array}
	{c} |\uparrow\uparrow\ \rangle \\ |\uparrow\downarrow\ \rangle \\ |\downarrow\uparrow\ \rangle \\ |\downarrow\downarrow\ \rangle \end{array}\right(
	& \begin{array}{c} \frac{1-p}{4} \\ 0 \\ 0\\ 0 \end{array}
	& \begin{array}{c} 0 \\ \frac{1+p}{4} \\ \frac{p}{2}\\ 0 \end{array}
	& \begin{array}{c} 0 \\ \frac{p}{2}\\ \frac{1+p}{4}\\ 0 \end{array}
	& \begin{array}{c} 0 \\ 0 \\ 0 \\ \frac{1-p}{4} \end{array}
	& \left)\begin{array}{c} \\ \\ \\  \\ \end{array}\right.
	\end{array}
	\end{split}
	\end{equation}
	By Eq.(\ref{relationF}) we could transform it according to another measurement $1x2x$ (both magnetic field in channel I and II are along x-axis) as:
	
	\begin{equation}
	\begin{split}
	\rho=\begin{array}{@{}r@{}c@{}c@{}c@{}c@{}l@{}}
	& |++\ \rangle & |+-\ \rangle & |-+\ \rangle & |--\  \rangle  \\
	\left.\begin{array}
	{c} |++\ \rangle \\ |+-\ \rangle \\ |-+\ \rangle \\ |--\ \rangle \end{array}\right(
	& \begin{array}{c} \frac{1+p}{4} \\ 0 \\ 0\\ -\frac{p}{2} \end{array}
	& \begin{array}{c} 0 \\ \frac{1-p}{4} \\ 0\\ 0 \end{array}
	& \begin{array}{c} 0 \\ 0\\ \frac{1-p}{4}\\ 0 \end{array}
	& \begin{array}{c} -\frac{p}{2} \\ 0 \\ 0 \\ \frac{1+p}{4} \end{array}
	& \left)\begin{array}{c} \\ \\ \\  \\ \end{array}\right.
	\end{array}
	\end{split}
	\end{equation}
	Then we could predict its spectrum under measurement $1x2x$ as:
	
	\begin{eqnarray}
	\begin{split}
	S(|\Psi_{s}\rangle\langle\Psi_{s}|, z, z, {\bf x_1}, {\bf x_2})= 
	\ &\frac{1+p}{4} S(|+\rangle\langle +|, x, {\bf x_1})S(|+\rangle\langle +|, x, {\bf x_2}) \\
	+&\frac{1-p}{4}S(|+\rangle\langle +|, z, {\bf x_1})S(|-\rangle\langle -|, z, {\bf x_2}) \\
	+&\frac{1-p}{4}S(|-\rangle\langle -|, z, {\bf x_1})S(|+\rangle\langle +|, z, {\bf x_2}) \\
	+&\frac{1+p}{4}S(|-\rangle\langle -|, z, {\bf x_1})S(|-\rangle\langle -|, z, {\bf x_2})
	\end{split}
	\end{eqnarray}
	
	Similarly, we would obtain the spectrum under other measurements. The simulation results of the Werner state are shown in Fig.(10), where we set $p=0.6$.
	\begin{figure}
		\begin{center}
			\fbox{\includegraphics[width=15cm,clip]{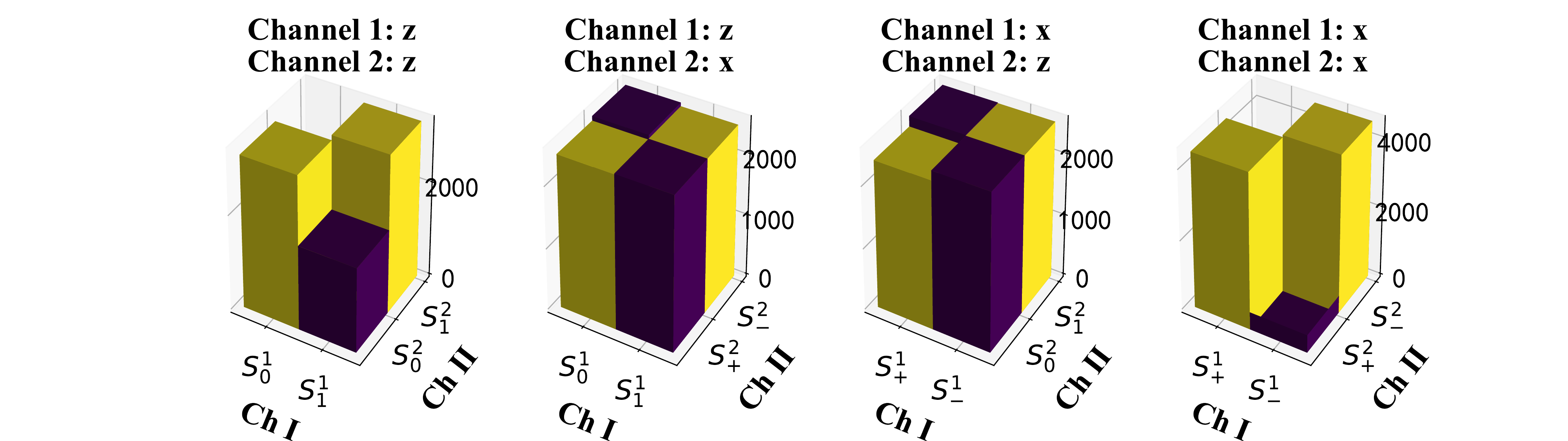}}
			\caption{{\bf Simulation result (Werner state, $\rho = 0.6$)}\\
				Left1 : Both the magnetic field in SG apparatus of channel 1 and 2 are along x-axis. Left2: Magnetic field in SG apparatus of channel 1 is along z-axis, and 2 is along x-axis. Right 2: Magnetic field in SG apparatus of channel 1 is along x-axis, and 2 is along z-axis. Right 1: Both the magnetic field in SG apparatus of channel 1 and 2 are along z-axis.).}	
		\end{center} 
	\end{figure}

\end{document}